\documentclass[a4,12pt]{article}

\usepackage{amsmath,amssymb,amsfonts,graphics,graphicx,amscd,amsfonts,epsf,epsfig,color}
\makeatletter
\@addtoreset{equation}{section}
\renewcommand{\theequation}{\thesection.\@arabic\c@equation}
\makeatother

\makeatletter
\renewcommand\appendix{\par%\newpage
  \setcounter{section}{0}%
  \setcounter{subsection}{0}%
  \gdef\thesection{Appendix \@Alph\c@section }
  \renewcommand{\theequation}
  {\Alph{section}.\arabic{equation}}
}
\makeatother

\newcommand{\ba}{\begin{eqnarray}}
\newcommand{\ea}{\end{eqnarray}}

\newcommand{\nt}{\notag\\}

\newcommand{\vect}[2]{\vec{#1}^{\;\!#2}}

\date{}
\begin{document}

\begin{titlepage}

\begin{center}
  {\LARGE
  Microstates of D1-D5(-P) black holes \\[+4pt] as interacting D-branes
  }
\end{center}
\vspace{0.2cm}
\baselineskip 18pt 
\renewcommand{\thefootnote}{\fnsymbol{footnote}}

\begin{center}
Takeshi {\sc Morita}${}^{a}$\footnote{%
E-mail address: morita.takeshi@shizuoka.ac.jp
}, 
Shotaro S{\sc hiba}${}^{b}$\footnote{
E-mail address: sshiba@cc.kyoto-su.ac.jp}

\renewcommand{\thefootnote}{\arabic{footnote}}
\setcounter{footnote}{0}

\vspace{0.4cm}

{\small\it

${}^a$ Department of Physics,
Shizuoka University, \\
836 Ohya, Suruga-ku, Shizuoka 422-8529, Japan

${}^b$ Maskawa Institute for Science and Culture, Kyoto Sangyo University, \\
Kamigamo-Motoyama, Kita-ku, Kyoto 603-8555, Japan.

}

\end{center}

%\newpage

\vspace{1.5cm}

\begin{abstract}
In our previous study \cite{Morita:2013wfa}, we figured out that the thermodynamics of the near extremal black $p$-branes can be explained as the collective motions of gravitationally interacting elementary $p$-branes (the $p$-soup proposal).
We test this proposal in the near-extremal D1-D5 and D1-D5-P black holes and show that their thermodynamics also can be explained in a similar fashion, i.e. via the collective motions of the interacting elementary D1-branes and D5-branes (and waves).
It may imply that the microscopic origins of these intersecting black branes and the black $p$-brane are explained in the unified picture.
We also argue the relation between the $p$-soup proposal and the conformal field theory calculations of the D1-D5(-P) black holes in superstring theory.

\end{abstract}
%\\

%\\

\end{titlepage}

%%%%%%%%%%%%%%%%%%%%%%%%%%%%%%%%%%%%%%%%%%%%%%%%%%%%%%%%%%%%%%%%%%%%%%
%%%%%%%%%%%%%%%%%%%%%%%%%%%%%%%%%%%%%%%%%%%%%%%%%%%%%%%%%%%%%%%%%%%%%%
%%%%%%%%%%%%%%%%%%%%%%%%%%%%%%%%%%%%%%%%%%%%%%%%%%%%%%%%%%%%%%%%%%%%%%
\newpage
\baselineskip 18pt
%%%%%%%%%%%%%%%%%%%%%%%%%%%%%%%%%%%%%%%%%%%%%%%%%%%%%%%%%%%%%%%%%%%%%%
%%%%%%%%%%%%%%%%%%%%%%%%%%%%%%%%%%%%%%%%%%%%%%%%%%%%%%%%%%%%%%%%%%%%%%
%%%%%%%%%%%%%%%%%%%%%%%%%%%%%%%%%%%%%%%%%%%%%%%%%%%%%%%%%%%%%%%%%%%%%%
\section{Introduction}
One of the most remarkable achievements in the superstring theory is the microscopic computations of several classes of the (near-)extremal black hole entropies initiated by the work of Strominger and Vafa \cite{Strominger:1996sh}. (See reviews \cite{Aharony:1999ti, David:2002wn, Mandal:2010cj, Dabholkar:2012zz}.)
These results provide the microscopic descriptions of these black holes, and they are the strong evidence that the superstring theory works as the quantum gravity at the non-perturbative level.
However these studies have been mainly developed in the intersecting black branes, especially in the D1-D5 system \cite{Callan:1996dv}, and it is the outstanding problem whether string theory can explain the thermodynamics of other black holes.

Recently, it has been shown that the thermodynamics of the near-extremal black $p$-branes in supergravity may be explained by an effective theory of gravitationally interacting elementary $p$-branes \cite{Morita:2013wfa}. (Related studies have been done in \cite{Horowitz:1997fr, Li:1997iz, Banks:1997tn, Li:1998ci, Smilga:2008bt, Wiseman:2013cda, Morita:2013wla}.)
The elementary branes may compose a bound state at low energy due to the strong gravitational force, and, by using the virial theorem, we can estimate the free energy of the bound state as functions of physical parameters: gravitational coupling, brane tension, the number of the elementary branes and temperature.
Then their dependence on the parameters agrees with those of the corresponding black brane.
Also the size of the bound state agrees with the size of the event horizon of the black brane.
(Interestingly we are naturally able to reproduce $\pi$ dependence too.)
We call this proposal `warm $p$-soup' \cite{Morita:2013wfa}, since the bound state is strongly coupled.

The $p$-soup proposal works for general near extremal black $p$-branes including branes in the superstring theory, e.g. D$p$, M$p$, F1 and NS5-branes \cite{Morita:2013wfa,Wiseman:2013cda, Morita:2013wla,Morita:2014}.
Then it is natural to ask whether the $p$-soup proposal can explain the intersecting black branes.
In this letter, we will study the near-extremal D1-D5(-P) system and show that indeed the $p$-soup proposal may work.
It may imply that the microscopic origins of the intersecting black branes and black $p$-branes are explained in the unified way.
We will also compare this result and the conformal field theory calculations of the D1-D5(-P) system in string theory \cite{Callan:1996dv}.

\section{D1-D5 system}

To study the D1-D5 black hole, we consider IIB superstring theory compactified on $ S^1 \times T^4$ and 
put $Q_1$ D1-branes and $Q_5$ D5-branes winding on $S^1$ and $S^1 \times T^4$ respectively.
(See Table \ref{table-D1D5P}.)
We take the size of $T^4$ small so that the D1-branes are uniformly smeared over there.

\begin{table}
\begin{eqnarray*}
\begin{array}{l|c|cccc|c|c|}
& t & 1 & 2& 3 &4 & (5) & T^4 \\
\hline
Q_1~\text{D1-brane} & - & &&& & - &   \\ \hline
Q_5~\text{D5-brane}& - & &&& & - & -   \\ \hline
N~\text{P (wave)} & - & &&& & - &   
\\ \hline
\end{array}
\end{eqnarray*}
\caption{The brane configuration of the D1-D5-P system.
The configuration of D1-D5 system is the same one with $N=0$.
We take $x^5$ as the $S^1$ coordinate with the period $2\pi R$.
}
\label{table-D1D5P}
\end{table}

If the branes are static, this configuration is BPS and no forces work.
However if they are moving, the interactions arise.
Our proposal is that these interactions confine the branes 
to a finite region and they compose a bound state, and
 this bound state explains the thermodynamics of the D1-D5 black hole.
To see this, we estimate the low energy effective action of this interacting brane system.
We assume that the branes are well separated and the gravitational interactions dominate.
Although we can calculate these interactions  between the branes from the IIB supergravity in a similar manner to \cite{Okawa:1998pz}, we use a shortcut \cite{Morita:2013wfa}.
We can read off the interactions from the probe D1-brane action in the extremal D1-D5 brane background \cite{David:2002wn},
\begin{align}
&S_{\text{D1}}^{\text{probe}}
= -m_1 \int dt \left( \frac{1}{H_1} \sqrt{1-H_1 H_5 \vect{v}{2}} - \left(\frac{1}{H_1} -1 \right) \right),  \label{probe-D1} \nt
&H_1=1+\frac{r_1^2}{\vect{r}{2}}, \quad H_5=1+\frac{r_5^2}{\vect{r}{2}}, \quad
r_1^2= \frac{ 4 m_1 G_5 Q_1}{\pi}, \quad 
r_5^2= \frac{4 m_5 G_5 Q_5}{\pi}.
\end{align}
Here we have taken the radius of the $S^1$ as $R$, and assumed that $R$ is small and the probe D1-brane depends on the time $t$ only \footnote{
The small $R$ assumption is not essential in the following calculations, and
we apply it only to make the equations simpler.
Note that if $R$ becomes very small, a phase transition related to the Gregory-Laflamme 
transition will occur \cite{Martinec:1999sa} at the point (\ref{T_GL}).
}. 
$\vec{r}$ and $\vec{v} \equiv \partial_t \vec{r}$ are the position and the velocity of the D1-brane in the (non-compact) 4 dimensional space.
$G_5 \equiv  4 \pi^5 g_s^2 \alpha'^4 /V_4 R  $ is the 5-dimensional Newton constant where
 $g_s$ and $\alpha'$ are the string coupling and the Regge parameter and $V_4$ is the volume of $T^4$.
$m_1$ and $m_5$ are masses of single D1 and D5-brane defined by
\begin{align}
m_1 \equiv \frac{ R}{g_s \alpha'} , \quad
m_5 \equiv \frac{ RV_4}{(2\pi)^4g_s \alpha'^3}.
\end{align}

We assume that the velocity $|\vec{v}|$ is small at low energy and expand the probe action as
\begin{align}
&S_{\text{D1}}^{\text{probe}}
= \int dt \left[ -m_1 + \frac{m_1}{2} \vec{v}^2 + \frac{m_1}{2} \frac{r^2_5}{r^2} \vec{v}^2 
+ \frac{m_1}{8} \vec{v}^4 + \frac{m_1}{8} \frac{r_1^2}{r^2} \vec{v}^4  + \frac{m_1}{8} \frac{r_1^2 r_5^4}{r^6} \vec{v}^4 + \cdots
 \right].
 \label{probe-D1-expansion}
\end{align}
Similarly from the probe D5-brane action in the same background, we obtain 
\begin{align}
&S_{\text{D5}}^{\text{probe}}
= \int dt \left[ -m_5 + \frac{m_5}{2} \vec{v}^2 + \frac{m_5}{2} \frac{r^2_1}{r^2} \vec{v}^2 
+ \frac{m_5}{8} \vec{v}^4 + \frac{m_5}{8} \frac{r_5^2}{r^2} \vec{v}^4  + \frac{m_5}{8} \frac{r_5^2 r_1^4}{r^6} \vec{v}^4 + \cdots
 \right].
 \label{probe-D5-expansion}
\end{align}
From these expansions, we can speculate the effective action of the separated $Q_1$ D1-branes and $Q_5$ D5-branes.
The first terms of these expansions are just the rest masses, 
and we will omit them in the following discussion. 
The second and fourth terms are the non-relativistic kinetic terms and their relativistic corrections.
Thus the effective Lagrangian of the branes must include 
\begin{align}
\sum_{i=1}^{Q_1} \left( \frac{m_1}{2} \vec{v_i}^2+ \frac{m_1}{8} (\vec{v_i}^2)^2+ \cdots \right)
+\sum_{i=1}^{Q_5} \left( \frac{m_5}{2} \vec{v_i}^2+ \frac{m_5}{8} (\vec{v_i}^2)^2 + \cdots \right).
\label{eff-kin}
\end{align}

The interactions between the branes can be read from the other terms of the expansions  (\ref{probe-D1-expansion}) and (\ref{probe-D5-expansion}).
The third term of (\ref{probe-D1-expansion}) is the two-body interaction between the probe D1-brane and the background D1-D5 geometry, and, since this interaction is independent of the D1-charge of the background geometry, we can read off the interaction between the single D1-brane and the single D5-brane by replacing $Q_5 \to 1$.
Then we obtain the two-body interactions between the separated $Q_1$ D1-branes and $Q_5$ D5-branes as
\begin{align}
L_1 \equiv \sum_{i=1}^{Q_1} \sum_{j=1}^{Q_5} \frac{ 2G_5 m_1 m_5}{\pi} \frac{\vect{v}{2}_{ij}}{\vect{r}{2}_{ij}}.
\label{eff-2-body-D1-D5}
\end{align}
Here $\vec{r}_{ij}$ and $\vec{v}_{ij}$ denote the relative position and relative velocity of the $i$-th and $j$-th branes. 
This interaction is consistent with the third term of the expansion (\ref{probe-D5-expansion}).
We define this term as $L_1$.
Similarly we can read off the two-body interactions between D1-branes and D5-branes from the fifth terms of (\ref{probe-D1-expansion}) and (\ref{probe-D5-expansion}), respectively, and obtain the interaction terms 
\begin{align}
\sum_{i\neq j}^{Q_1}  \frac{ G_5 m_1^2}{2\pi} \frac{\vect{v}{4}_{ij}}{\vect{r}{2}_{ij}}
+
\sum_{i\neq j}^{Q_5}  \frac{ G_5 m_5^2}{2\pi} \frac{\vect{v}{4}_{ij}}{\vect{r}{2}_{ij}}.
\label{eff-2-body}
\end{align}
Note that the power of $\vec{v}_{ij}$ of these interactions are higher than that of $L_1$, and
it implies that $L_1$ would dominate in the low energy regime where $|\vec{v}|$ would be small.

The last terms in (\ref{probe-D1-expansion}) and (\ref{probe-D5-expansion}) are proportional to $m_1^2 m_5^2 G_5^3$ indicating  three graviton (and RR-gauge and dilaton) exchange interactions among two D1-branes and two D5-branes,
\begin{align}
L_2 \equiv
\sum_{i=1}^{Q_1} \sum_{j=1}^{Q_1} \sum_{k=1}^{Q_5} \sum_{l=1}^{Q_5}
\frac{G_5^3  m_1^2 m_5^2}{\pi^3} \left(  \frac{\vect{v}{4}_{ij}}{\vect{r}{2}_{ij} \vect{r}{2}_{ik} \vect{r}{2}_{il}  }+ \cdots \right).
\label{eff-4-body}
\end{align}
We define this term as $L_2$.
The precise velocity dependences of these interactions cannot be determined from the probe actions, and we need to solve the multi-body problem as in \cite{Okawa:1998pz}.
However we will consider an order estimate for the thermodynamics of this system, and the precise expressions are not necessary and we leave this issue for future works.
In the same way, we can speculate other interactions from the terms which are not explicitly written in the expansions (\ref{probe-D1-expansion}) and (\ref{probe-D5-expansion}).

By combining the terms (\ref{eff-kin}) -- (\ref{eff-4-body}), we obtain the effective action for the interacting $Q_1$ D1-brane and $Q_5$ D5-brane systems:
\begin{align}
L_{D1D5}=&\sum_{i=1}^{Q_1} \left( \frac{m_1}{2} \vec{v_i}^2+ \frac{m_1}{8} (\vec{v_i}^2)^2+ \cdots \right)
+\sum_{i=1}^{Q_5} \left( \frac{m_5}{2} \vec{v_i}^2+ \frac{m_5}{8} (\vec{v_i}^2)^2 + \cdots \right) \nonumber \\
&+ \sum_{i=1}^{Q_1} \sum_{j=1}^{Q_5} \frac{ 2G_5 m_1 m_5}{\pi} \frac{\vect{v}{2}_{ij}}{\vect{r}{2}_{ij}}+
\sum_{i\neq j}^{Q_1}  \frac{ G_5 m_1^2}{2\pi} \frac{\vect{v}{4}_{ij}}{\vect{r}{2}_{ij}}
+
\sum_{i\neq j}^{Q_5}  \frac{ G_5 m_5^2}{2\pi} \frac{\vect{v}{4}_{ij}}{\vect{r}{2}_{ij}} \nonumber \\ 
&+\sum_{i=1}^{Q_1} \sum_{j=1}^{Q_1} \sum_{k=1}^{Q_5} \sum_{l=1}^{Q_5}
\frac{G_5^3  m_1^2 m_5^2}{\pi^3} \left(  \frac{\vect{v}{4}_{ij}}{\vect{r}{2}_{ij} \vect{r}{2}_{ik} \vect{r}{2}_{il}  }+ \cdots \right)+ \cdots .
\label{eff-D1D5}
\end{align}
The Lagrangian also have other terms arising from the expansions (\ref{probe-D1-expansion}) and (\ref{probe-D5-expansion}) but we will consider them later.
From now, we estimate the dynamics of this system by using the virial theorem.
We first assume that the branes are confined due to the interactions, and the branes satisfy
\begin{align}
\vec{v}_{ij} \sim  v, \qquad \vec{ r}_{ij} \sim r.
\label{assumption-scales}
\end{align}
Here $v$ and $r$ are the characteristic scales of the velocity and position of the branes in the bound state which do not depend on the species of the branes.
(Note that since the masses of the D1-brane and D5-brane are different, we naively expect that  these scales should depend on the species of the branes.
However we will soon see that it does not occur in the bound state.)
Then we can estimate the scales of the terms in the effective Lagrangian (\ref{eff-D1D5}) as
\begin{align}
L \sim & Q_1 m_1 v^2+ Q_1 m_1 v^4+ Q_5 m_5 v^2+ Q_5 m_5 v^4  \nonumber \\
& +\frac{ G_5 Q_1 Q_5 m_1 m_5}{\pi} \frac{v^2}{r^2} + \frac{ G_5 Q_1^2 m_1^2}{\pi} \frac{v^4}{r^2}
 + \frac{ G_5 Q_1^2 m_1^2}{\pi} \frac{v^4}{r^2} \nonumber \\
&+ \frac{G_5^3 Q_1^2 Q^2_5 m_1^2 m_5^2}{\pi^3} \frac{v^4}{r^6} + \cdots.
\label{estimate-D1D5}
\end{align}
where the ordering of the terms is the same as (\ref{eff-D1D5}).
`$\sim$' in this article denotes equality not only including dependence on physical parameters but also including all factors of $\pi$.
Here we consider which terms in (\ref{estimate-D1D5}) dominate at the low energy where $v$ would be small ($v \ll 1 $). 
In the second line of (\ref{estimate-D1D5}), the first term which is from $L_1$ (\ref{eff-2-body-D1-D5}) would dominate \footnote{If the numbers of the branes $Q_1$ and $Q_5$ are quite different, e.g. $Q_5 \gg Q_1$, we can ignore the another species of the branes and the results would be changed. }.
Suppose that this term is balanced to the term in the third line  which is from $L_2$ (\ref{eff-4-body}) due to the virial theorem, 
we obtain the relation between $v$ and $r$ as
\begin{align}
\frac{ G_5 Q_1 Q_5 m_1 m_5}{\pi} \frac{v^2}{r^2}
 \sim 
\frac{G_5^3 Q_1^2 Q^2_5 m_1^2 m_5^2}{\pi^3} \frac{v^4}{r^6} \quad \Longrightarrow \quad  
v^2 \sim \frac{\pi^2 r^4}{Q_1 Q_5 m_1 m_5 G_5^2} \left( \sim \frac{r^4}{r^2_1r^2_5} \right).
\label{scale-v-r}
\end{align}
Thus $r^2 \ll r_1^2, r_5^2$ would be satisfied  at low energy $v \ll 1 $.
We will later see that the limit $r^2 \ll r_1^2, r_5^2$ corresponds to the near extremal limit in gravity.

Under the the scaling relation (\ref{scale-v-r}), the terms in the Lagrangian (\ref{estimate-D1D5}) scale as,
\begin{align}
L \sim & \frac{ \pi r^2}{G_5} \frac{r^2}{r_5^2} + \frac{\pi  r^2}{G_5} \frac{r^2}{r_5^2} \frac{r^4}{r_1^2 r_5^2} +   \frac{\pi r^2}{G_5} \frac{r^2}{r_1^2}  + \frac{\pi  r^2}{G_5} \frac{r^2}{r_1^2} \frac{r^4}{r_1^2 r_5^2}  \nonumber \\
& +\frac{ \pi r^2}{G_5} +\frac{ \pi r^2}{G_5} \frac{r^4}{r_5^4}
+\frac{ \pi r^2}{G_5} \frac{r^4}{r_1^4}\nonumber \\
&+ \frac{ \pi r^2}{G_5} + \cdots.
\end{align}
We see that $L_1$ and $L_2$ scale as $ \pi r^2/G_5$, while the other terms earn the factors of $r^2/r_1^2$ and/or $r^2/r_5^2$ and are suppressed at low energy ($r^2 \ll r_1^2, r_5^2$).
Hence the scaling relation (\ref{scale-v-r}) is ensured self-consistently.
Note that the masses of the branes always appear as the combination $m_1 m_5$ in $L_1$ and $L_2$, and it ensures that the scales of the position $r$ and velocity $v$ are independent of the species of the branes as we assumed in (\ref{assumption-scales}).

So far we have considered the first several terms obtained from the expansions of the probe actions (\ref{probe-D1-expansion}) and (\ref{probe-D5-expansion}), and derived the scaling relation (\ref{scale-v-r}) at low energy via the virial theorem.
We now consider the contributions of the higher order terms in these expansions.
Since $r^2 \ll r_1^2, r^2_5 $ would be satisfied at low energy, we apply this approximation to the probe D1-brane action (\ref{probe-D1}) and expand it as
\begin{align}
S_{\text{D1}}^{\text{probe}}
= & \int dt \Biggl( \frac{2G_5 m_1 m_5 Q_5}{ \pi \vect{r}{2}} \vect{v}{2} 
 + \frac{8 G_5 m^2_1 Q_1 (G_5 m_5 Q_5)^2}{\pi^3 \vect{r}{6}} \vect{v}{4}  + \sum_{n=3}^\infty L_n^{\text{probe}} \Biggr), 
\label{probe-D1-expand}  \\
 L_n^{\text{probe}}= &\frac{2^{3n-2}(2n-3)!!}{n!}
\frac{ G_5 m_1 m_5 Q_5}{\pi \vect{r}2} 
\left( \frac{G_5^2 m_5 Q_5 m_1 Q_1}{\pi^2 \vect{r}4} \right)^{n-1} \vect{v}{2n}.
\label{multi-graviton} 
\end{align}
Here $L_n^{\text{probe}}$ describes the $2n-1$ graviton exchange interaction.
Through the similar speculations to the derivations of the interactions (\ref{eff-2-body-D1-D5}) -- (\ref{eff-4-body}), we estimate the effective action of separated $Q_1$ D1-branes and $Q_5$ D5-branes for $r^2 \ll r_1^2, r^2_5 $, as
\begin{align}
&S_{\text{D1D5}}= \int dt \sum_{n=1}^\infty L_n, 
 \label{moduli-D1D5} \\
L_{n} \sim & \sum_{i_1, \dots, i_n}^{Q_1}\sum_{j_1, \dots, j_n}^{Q_5} %\nonumber \\
\left( G_5^{2n-1} \frac{m_1^n m_5^n }{\pi^{2n-1}} \prod_{k=2}^n \prod_{l=1}^n
\frac{ 1 }{  \vect{r}{2}_{i_1 i_k} \vect{r}{2}_{i_1 j_l} } \vect{v}{2n} + \cdots \right),
\label{moduli-multi-graviton} 
\end{align}
The interaction $L_n$ represents the $2n-1$ graviton exchange among $n$ D1-branes and $n$ D5-branes.
Again the precise numerical coefficients for these interactions cannot be determined from the probe actions but it is not a matter for our purpose.

We can see that at the scaling (\ref{scale-v-r}) which was derived via the virial theorem $L_1 \sim L_2 $, all the other interactions $L_n$ also become the same order.
It means that the branes are strongly coupled in the bound state. For this reason, we called such a bound state as `warm $p$-soup' in Ref.~\cite{Morita:2013wfa}.

From now, we evaluate the thermodynamical quantities of the bound state.
By substituting the relation (\ref{scale-v-r}) to the Lagrangian $\sum L_n \sim L_1$, we estimate the free energy of the system as
\begin{align}
F \sim L_1 \sim  \frac{\pi r^2}{G_5}.
\label{free-energy-D1D5} 
\end{align}
Here we consider temperature dependence.
If the bound state is thermalized, we treat $\vec{r}_i$ as a thermal field (particle) and expand $\vec{r}_i(t) = \sum_n \vec{r}_{i(n)} \exp\left(i \frac{2 \pi n}{\beta}t  \right) $. 
Hence we assume that the velocity $v = \partial_t r$ are characterized by the temperature of the system through
\begin{align}
v  \sim \pi T r.
\label{key-assumption}
\end{align}
Note that such a relation is not held generally if the system has a mass gap, but
there would be no mass gap in the interacting brane system as argued in Ref.~\cite{Morita:2013wfa}.
Then, from (\ref{scale-v-r}), we obtain the relation between the size of the bound state and the temperature
\begin{align}
r \sim T  G_5 \sqrt{ Q_1 Q_5 m_1 m_5}. \label{scale-r-T}
\end{align}
By substituting this relation into the free energy (\ref{free-energy-D1D5}), 
we estimate the entropy of the bound state as
\begin{align}
&S_{\text{entropy}} = - \frac{\partial F}{\partial T}  \sim \pi m_1 m_5 Q_1 Q_5 G_5 T.
\label{entropy-D1D5}  
\end{align}

We compare the obtained quantities with the D1-D5 black hole \cite{Martinec:1999sa}.
In the near extremal regime, the black hole thermodynamics tells us, 
\begin{align}
F&=-\frac{\pi r_H^2}{8 G_5} ,
\label{free-energy-D1D5-GR} \\
S_{\text{entropy}}&= 16 \pi m_1 m_5 G_5 Q_1 Q_5 T, \label{entropy-D1D5-GR} \\
r_H&= 8 G_5 T \sqrt{m_1 m_5 Q_1 Q_5}.
\label{eq-tmp-horizon}
\end{align}
Here $r_H$ is the location of the horizon.
Therefore, if we identify the size of the bound state $r$ with the horizon $r_H$, our result (\ref{free-energy-D1D5}), (\ref{scale-r-T}) and (\ref{entropy-D1D5}) reproduce the parameter dependences of the black hole thermodynamics including $\pi$.
($r_H$ depends on the coordinate and we have argued what coordinate is natural in \cite{Morita:2013wfa}.)
This agreement may indicate that 
the interacting D1- and D5-branes described by the effective action (\ref{moduli-D1D5}) provide
the microscopic origin of the D1-D5 black hole thermodynamics.
Interestingly the free energy (\ref{free-energy-D1D5-GR}) has been reproduced without imposing the assumption (\ref{key-assumption}) about temperature, as in (\ref{free-energy-D1D5}).

We should emphasize that in order to derive the thermodynamical quantities from the effective theory (\ref{moduli-D1D5}) we have employed only the natural assumptions 
commonly used in interacting systems and the additional one (\ref{key-assumption}) about temperature which may be a characteristic property of  the branes at low energy \cite{Morita:2013wfa}.

Finally we comment on the assumption $r^2 \ll r_1^2, r_5^2$ which we have used when we consider the effective action (\ref{moduli-D1D5}).
At the scale (\ref{scale-r-T}), this relation becomes $T \ll 1/r_1, 1/r_5$ and this is the near extremal limit in supergravity \cite{David:2002wn}.
Thus our analysis is valid when we consider the near extremal black holes.
Moreover, according to supergravity, a phase transition related to the 
Gregory-Laflamme transition along the $S^1$ occurs around $r_H \sim \alpha'/R$ \cite{Martinec:1999sa}, hence through (\ref{eq-tmp-horizon})\footnote{Below this temperature, we should take a T-duality along $S^1$, which maps the D1- and D5-branes to D0- and D4-branes respectively and go to the IIA flame.
There the stable solution in supergravity is the D0-D4 black hole which is localized on the $S^1$.
We can reproduce the thermodynamical quantities of this black hole by considering  the interacting D0 and D4-brane model similar to the D1-D5 system \cite{wip}.
If temperature is below $T \sim \frac{g_s \alpha'^2}{R V_4}$, another phase transition occurs and the BPS matrix string describes the system \cite{Martinec:1999sa}.
\label{ftnt-GL}},
\ba\label{T_GL} 
T_{\text{GL}} \sim \frac{1}{g_s\alpha'R}\sqrt{\frac{V_4}{Q_1Q_5}}.
\ea
Therefore our results may be valid in the region $T_{GL}< T \ll  1/r_1, 1/r_5$.

%\vspace{5mm}
\section{D1-D5-P system}
We apply the similar analysis to the D1-D5-P system in this section.
We consider the same brane configuration to the D1-D5 system but now add momentum $N/R$ along $S^1$.
(See Table \ref{table-D1D5P}.)

To derive the effective theory of this system, we consider the gravitational interactions among the branes and the gravitational waves which carry the momentum $1/R$ along $S^1$.
First we look at the probe D1-brane in the extremal D1-D5-P background \cite{David:2002wn}
\begin{align}
S_{\text{D1}}^{\text{probe}}=& -\frac{R}{ g_s \alpha' } \int dt \left( \frac{1}{H_1} \sqrt{1-H_1 H_5 H_p \vect{v}2} - \left(\frac{1}{H_1} -1 \right) \right),  \label{probe-D1-3charge} \nt
&H_p = 1+ \frac{r_p^2}{\vect{r}2}, \quad r_p^2 = \frac{4 G_5 N}{\pi R}.
\end{align}
Then, by repeating the arguments in the previous section, we can estimate the effective theory for the branes and waves.
At low energy, $r \ll r_1,r_5, r_p$ would be satisfied at the bound state similar to the D1-D5 system, and we estimate the effective action as
 \begin{align}
&S_{\text{D1D5P}}= \int dt \sum_{n=1}^\infty L_n, \quad
 \label{moduli-D1D5P} \nonumber \\
& L_1 \sim \sum_{i=1}^{Q_1} \sum_{j=1}^{Q_5} \sum_{k=1}^{N}  
\frac{G_5^2 m_1 m_5 }{\pi^2 R} \frac{\vect{v}{2}_{ij}}{\vect{r}{2}_{ij}\vect{r}{2}_{ik}}+ \cdots, \nonumber \\
& L_2\sim
\sum_{i,j=1}^{Q_1} \sum_{k,l=1}^{Q_5} \sum_{m,n=1}^{N}  
\frac{G_5^5 m_1^2 m_5^2 }{\pi^5 R^2} \frac{\vect{v}{4}_{ij}}{ \vect{r}{2}_{ij} \vect{r}{2}_{ik} \vect{r}{2}_{il} \vect{r}{2}_{im} \vect{r}{2}_{in} } + \cdots, \nt
&  L_{n} \sim \sum_{i_1, \dots, i_n}^{Q_1} \sum_{j_1, \dots, j_n}^{Q_5}  \sum_{k_1, \dots, k_n}^{N}  \left( \frac{G_5^{3n-1}  m_1^n m_5^n }{\pi^{3n-1} R^n} \prod_{a=2}^n \prod_{b=1}^n \prod_{c=1}^n 
\frac{ 1}{ \vect{r}{2}_{i_1 i_a} \vect{r}{2}_{i_1 j_b} \vect{r}{2}_{i_1 k_c} } \vect{v}{2n}+  \cdots \right).
\end{align}
$L_n$ describes the interactions among $n$ D1-branes, $n$ D5-branes and $n$ waves through the exchanges of the $3n-1$ gravitons.
Although these schematic expressions can be predicted from the probe action, we need to solve the multi-body problem in the supergravity to determine the precise expressions.
Note that the interactions shown in the D1-D5 action (\ref{moduli-D1D5}) exist in this system too, but they are subdominant in the limit $r \ll r_1,r_5, r_p$ and they have been omitted here.

We estimate the free energy of this system by imposing the same assumptions to the D1-D5 case (\ref{assumption-scales}) and applying the virial theorem $L_1 \sim L_2$ to the effective action (\ref{moduli-D1D5P}). Then we obtain
\begin{align}
v^2  \sim \frac{\pi^2 r^6}{Q_1 Q_5 N G_5^2}, \qquad
F  \sim  L_1 \sim  \frac{\pi r^2}{G_5}.
\end{align}
In this derivation, we have used the relation $m_1 m_5/R = \pi/4G_5$.
To consider the temperature dependence, 
we further assume the relation (\ref{key-assumption}) and obtain
\begin{align}
 r^2 &  \sim  G_5  T  \sqrt{  NQ_1Q_5},
 \label{eq-scale-D1D5P}
  \\
 S_{\text{entropy}} & = -\frac{\partial F}{\partial T} \sim \pi \sqrt{ NQ_1Q_5  }.
\end{align}

Here we compare these results with the D1-D5-P black hole in the near extremal regime ($r_H \ll r_1,r_5, r_p$) \cite{David:2002wn}
\begin{align}
F=&-\frac{\pi r_H^2}{4G_5}, \\
 r_H^2  =& 8 G_5 T \sqrt{  NQ_1Q_5 }, \\
S=&
2\pi   \sqrt{ Q_1 Q_5 N}.
\end{align}
Therefore by identifying $r \sim r_H $, these results are consistent with our results including the $\pi$ dependence.

Our analysis is valid as far as $r^2 \ll r_1^2, r_5^2, r_p^2$ which correspond to the near extremal limit in the supergravity \cite{David:2002wn}.
Contrast to the D1-D5 case, no phase transition would occur at low temperature, and thus our  calculations may be valid until zero temperature.

\section{Discussions}
We have studied the  effective theories of the gravitationally interacting elementary branes (\ref{moduli-D1D5}) and (\ref{moduli-D1D5P}), and showed that these theories, with the natural assumptions, explain the D1-D5(-P) black hole thermodynamics in the near extremal regime.
It is remarkable that such a simple model describes the black hole microstates.
In \cite{Morita:2013wfa}, we have applied the same analysis to the interacting $p$-brane system, and reproduced the black $p$-brane thermodynamics (the $p$-soup proposal).
These successes suggest that we can understand the microstates of these distinct types of black holes in the unified fashion.

The D1-D5(-P) system has been also investigated through the conformal field theory (CFT) which appears at the IR fixed point of the Higgs branch of the gauge theory on the branes, and it reproduces the black hole thermodynamics exactly \cite{David:2002wn, Callan:1996dv}.
Here we compare this computation and ours.

\vspace{3mm}
\noindent
1. In the CFT calculation, we assume that the branes are coincident in the transverse 4 dimensional space (Higgs branch) \footnote{Ref.\,\cite{Seiberg:1999xz} argued the connection between the coincident branes and separated branes.}.
Since no force works between the branes at zero temperature,
in order to retain the branes coincident, we need to turn on the NS-NS $B$-field.
Thus the CFT calculation is done at the distinct point in the moduli space from the black holes ($B=0$), and the agreement to the gravity would be due to the non-renormalization theorem \cite{David:2002wn}.
Therefore the CFT calculation would not work for the quantities which are not protected by the supersymmetry or if the black hole is far from the near extremal regime.
 This point is different from the $p$-soup proposal, where we treat the separated D-branes 
with $B=0$.
This means that the $p$-soup proposal may describe the system at the same point in the moduli space as the black holes. 
In this sense we may regard the $p$-soup proposal as a direct description of the black hole microstates and it might capture even the properties of the black holes which are not protected by the non-renormalization theorem.

\vspace{3mm}
\noindent
2. In the $p$-soup proposal, the effective theories of the branes (\ref{moduli-D1D5}) and (\ref{moduli-D1D5P}) are obtained from the classical supergravity whereas the CFT is obtained from the gauge theory. 
Thus the CFT relies on the gauge/gravity correspondence in   superstring theory which is related to the UV structures of gravity. 
On the other hand, the $p$-soup proposal might work independently from the UV structure and it may suggest that the black hole thermodynamics may be explained via low energy properties of the supergravity. \footnote{The $p$-soup proposal does not work where the gravity description does break down.
 For example, the effective theory (\ref{eff-D1D5}) is not valid below $T_{\text{GL}}$ (\ref{T_GL}) in the D1-D5 system. (See footnote \ref{ftnt-GL}.)
In order to describe such circumstances, we would need the UV structures of gravity such as a T-dual of the theory or dual gauge theories.
In this sense our proposal does not provide a genuine microstate of quantum gravity.
Our claim is, however, that the statistical mechanics of gravitationally interacting branes % is enough to 
suffices for a microscopic explanation of the black hole thermodynamics
and that a genuine microstate of quantum gravity may not be necessary.  
}

\vspace{3mm}
In this way, these two microscopic descriptions of the black holes are quite different.
Although the exact computation in the $p$-soup proposal has not been done so far, 
the $p$-soup proposal has revealed the new aspects of the supergravity, and
we expect this proposal will play an important role to develop our understanding of black hole microstates.

%\vspace{3mm}
%\paragraph*{
\subsection*{Acknowledgements}
We would like to thank Andrew Hickling, Toby Wiseman and Benjamin Withers for useful comments through the collaboration. 
We would also like to thank Gautam Mandal and Yuji Okawa for important discussions.
T.M. is supported in part by National Science Foundation (NSF-PHY-1214341). 

%\appendix

%-----------------------------------------------------------------
%
%\bibliographystyle{jplain}
\bibliographystyle{utphys}
\bibliography{D1D5}

%
%-----------------------------------------------------------------

\end{document}